\documentclass[prd,preprint,superscriptaddress,nofootinbib]{revtex4}

\usepackage{axodraw}

\newlength{\intwidth}
\DeclareRobustCommand{\fpint}[2]
   {\mathop{%
      \text{%
        \settowidth{\intwidth}{$\int$}%
        \makebox[0pt][l]{\makebox[\intwidth]{$-$}}%
        $\int_{#1}^{#2}$}}}

\def\half{\textstyle{\frac{1}{2}}}

\begin{document}

\title{Four-Loop Cusp Anomalous Dimension From Obstructions}

\author{Freddy Cachazo}
\email{fcachazo@perimeterinstitute.ca}

\affiliation{Perimeter Institute for Theoretical Physics, Waterloo,
Ontario N2J 2W9, Canada}

\author{Marcus Spradlin}
\email{spradlin@het.brown.edu}

\affiliation{Brown University, Providence, Rhode Island 02912, USA}

\author{Anastasia Volovich}
\email{nastja@het.brown.edu}

\affiliation{Brown University, Providence, Rhode Island 02912, USA}

\begin{abstract}

We introduce a method for extracting the cusp anomalous dimension at
$L$ loops from four-gluon amplitudes in ${\cal N}=4$ Yang-Mills
without evaluating any integrals that depend on the kinematical
invariants. We show that the anomalous dimension only receives
contributions from the obstructions introduced in hep-th/0601031. We
illustrate this method by extracting the two- and three-loop
anomalous dimensions analytically and the four-loop one numerically.
The four-loop result was recently guessed to be $f^{(4)} = - \left(
4\zeta^3_2+24\zeta_2\zeta_4+50\zeta_6- 4(1+r)\zeta_3^2\right)$ with
$r=-2$ using integrability and string theory arguments in
hep-th/0610251. Simultaneously, $f^{(4)}$ was computed numerically
in hep-th/0610248 from the four-loop amplitude obtaining, with best
precision at the symmetric point $s=t$, $r=-2.028(36)$. Our
computation is manifestly $s/t$ independent
and improves the precision to $r=-2.00002(3)$,
providing strong evidence in favor of the conjecture.
The improvement is possible due to a large reduction in the number of
contributing terms, as well as a reduction in the number of
integration variables in each term.

\end{abstract}

\preprint{hep-th/0612309}
\preprint{BROWN-HET-1477}

\maketitle

\section{Introduction}

${\cal N}=4$ super Yang-Mills (MSYM) is a very fascinating theory.
Recently, important progress has been achieved in two different
but connected aspects of the theory. On the one hand, insights from
integrability have led to a proposal~\cite{Beisert:2006ib,Beisert:2006ez}
for all-loop quantities such as
the cusp anomalous dimension $f(g)$, which is the coefficient
of the logarithmic correction to the anomalous dimension of large
spin twist-2
operators~\cite{Korchemsky:1988si,Korchemsky:1992xv,
Gubser:2002tv,Kruczenski:2002fb}.
On the other hand,
in~\cite{Bern:2006ew} the integrand of the
planar four-loop four-gluon scattering amplitude was found,
and the amplitude was evaluated
analytically to order $1/\epsilon^4$ and numerically to order
$1/\epsilon^2$ in the dimensional regularization parameter
$\epsilon = 2 - D/2$.

The connection between these two remarkable results is that
the coefficient $f^{(L)}$ of $g^{2L}$ in the perturbative expansion of $f(g)$
controls part of the IR singular behavior of the $L$-loop scattering
amplitude~\cite{Magnea:1990zb,Catani:1998bh,Sterman:2002qn}.
Hence one can in principle extract that coefficient,
and thereby determine $f(g)$ order by order,
by calculating four-gluon scattering
amplitudes. However, in practice this is a very hard task. One
reason for this is that the IR singularity for an $L$-loop
amplitude begins at order $1/\epsilon^{2L}$,
while the contribution from $f(g)$ only appears in the
$1/\epsilon^2$ singularity. Therefore, the larger $L$, the higher
one has to climb into the $\epsilon$ expansion
before uncovering the desired
contribution from $f(g)$.

One of the results of~\cite{Beisert:2006ez} is a formula
for $f(g)$ whose expansion can be written in the form
\begin{equation}
\label{eq:cuspdimension} f(g) = 4g^2 -4\zeta_2 g^4 + \left(
4\zeta^2_2+12\zeta_4 \right) g^6 - \left(
4\zeta^3_2+24\zeta_2\zeta_4 + 50\zeta_6 -4(1+r)\zeta_3^2\right)
g^8+{\cal O}(g^{10}),
\end{equation}
where $r$ was conjectured
to have the value $r=-2$ and $g^2 = g^2_{\rm YM}N/(8\pi^2)$. This formula
refines an earlier guess in~\cite{Eden:2006rx} which had
the same form but $r=0$. The
non-zero value of $r$ is due to a nontrivial phase factor in the
spin chain S-matrix. Remarkably, the value $r=-2$ at four loops follows from a
simple conjecture for the all-loop phase factor also given
in~\cite{Beisert:2006ez} which has been shown to agree
well with the AdS/CFT correspondence~\cite{Benna:2006nd,Maldacena:2006rv}.

On the other hand, the numerical evaluation of the $1/\epsilon^2$
coefficient of the four-loop amplitude was carried out
in~\cite{Bern:2006ew}. The numerical calculation is done at
different kinematical points $x=t/s$ with results $(r,x)$ given by
$(-2.028(36),1)$, $(-2.059(36),2)$, $(-2.062(45),3)$,
$(-2.074(104),15)$, where the quantities in parentheses denote the
uncertainty in the last digits.

An unfortunate complication inherent in this approach is that the
$L$-loop cusp anomalous dimension is just a number, but the
$1/\epsilon^2$ coefficient of the $L$-loop scattering amplitude is,
in general, a very complicated function of the kinematic invariant
$x = t/s$. Isolating this single number is like picking a needle
out of a haystack. For example, the evaluation of the four-loop
amplitude using Mellin-Barnes
representations~\cite{Smirnov:1999gc,Tausk:1999vh,SmirnovBook,Czakon:2005rk}
(the current state
of the art) expresses the $1/\epsilon^2$ coefficient as a sum of
over 50,000 $x$-dependent terms\footnote{Each ``term''
here refers to an integral of a single rational expression
of $\Gamma$ functions and their derivatives.}.
Each of these terms is a multiple
integral which in general is evaluated numerically.

The fact that an appropriate combination of these terms conspires to
add up to an $x$-independent
number highlights an inefficiency in this calculational
approach.
This is manifest in the fact that the numerical precision
obtained by~\cite{Bern:2006ew} depends on the value of $x$ where the
evaluation is performed, as noted above, even though
the result of the calculation cannot depend on $x$. In terms of
$r$, the precision
reported by~\cite{Bern:2006ew}
varies from approximately $1.5\%$
at $x=1$ to $5\%$
at $x=15$.

In this paper we demonstrate that the cusp anomalous dimension may
be calculated in a manifestly $x$-independent manner, and with a
significantly smaller set
of terms with reduced number of integrations variables.
The starting point at four loops is the
formula~\cite{Magnea:1990zb,Catani:1998bh,Sterman:2002qn,Bern:2005iz}
\begin{eqnarray}
\label{eq:one}
M^{(4)}(x,\epsilon)&=& \frac{1}{4}\left( M^{(1)}(x,\epsilon)\right)^4
- \left( M^{(1)}(x,\epsilon)\right)^2 M^{(2)}(x,\epsilon)
+ M^{(1)}(x,\epsilon)M^{(3)}(x,\epsilon)\cr
&&\qquad + \frac{1}{2}\left(
M^{(2)}(x,\epsilon)\right)^2
+ \frac{1}{4} f^{(4)} M^{(1)}(x,4 \epsilon) + {\cal O}(1/\epsilon),
\end{eqnarray}
which relates the four-loop amplitude to a polynomial in lower-loop
amplitudes, together with a `correction' proportional to $f^{(4)}$.
Rather than calculating both sides of this equation as functions of
$x$, and then reading off $f^{(4)}$, as was done
in~\cite{Bern:2006ew}, we show that $f^{(4)}$ (and more generally
$f(g)$) only receives contributions from the `obstructions' discovered
in~\cite{Cachazo:2006mq}. Moreover, only the lowest order
obstructions contribute. It turns out that obstructions can be
systematically extracted from $x$-dependent integrals and the rest
can be thrown away term-by-term, eliminating the need to evaluate
all of them and then rely on non-trivial cancellations. We carry out
the relevant calculation analytically for the one-, two- and
three-loop amplitudes. At four loops we use Czakon's ${\tt MB}$
program~\cite{Czakon:2005rk} as a starting point and then implement
the systematic extraction of obstructions. Quite nicely, it turns
out that a large number of the integrals that come out can be
reduced by simple applications of Barnes lemmas and corollaries.
This process was also automated. After performing the remaining
integrations numerically, we obtain
\begin{equation}
f^{(4)} = -117.1789(2), \qquad {\rm or} \qquad
r = -2.00002(3),
\end{equation}
which is consistent with the prediction $r=-2$
within the precision of about $0.001\%$.

In fact this computation of the cusp anomalous dimension is a simple
application of a more general analysis of the structure of
multi-loop amplitudes we present in this paper.  Instead of working
with amplitudes as functions of $x$, we work with their Mellin
transforms to a new variable $y$. In $y$ space, any amplitude can be
uniquely expressed as a sum of two kinds of terms.  The first kind
are singularities, which take the form of $\delta$-functions or
derivatives of $\delta$-functions at $y=0$.  These singularities are
the obstructions introduced in~\cite{Cachazo:2006mq}; in $x$-space
they correspond to polynomials in $\log^2 x$. The remaining part of
the amplitude is smooth at $y=0$. We show that the obstructions in
$L$-loop amplitudes must {\it separately} satisfy any polynomial
relation of the form~(\ref{eq:one}) and that furthermore the cusp
anomalous dimension only receives contributions from the leading
singularity $\delta(y)$ at order $1/\epsilon^2$. To summarize:
\begin{quote}
Our prescription for computing the cusp anomalous dimension is to
 read off the numerical coefficient in front of $\delta(y)/\epsilon^2$
on both sides of the Mellin transform of equation~(\ref{eq:one}).
\end{quote}

This paper is organized as follows. In section 2 we review how the
cusp anomalous dimension
may be extracted from the infrared singularities of gluon scattering
amplitudes.
In section~3 we define obstructions in detail, discuss their most
important properties, and explain how they may be used to calculate
the
cusp anomalous dimension.
In section 4 we
illustrate the method at the two- and three-loop level exhibiting the
form of the obstructions explicitly. In section~5 we give some
details
on the computation of the four-loop cusp anomalous dimension from obstructions.
In section 6 we summarize our conclusions and list some possible
promising
directions for future work.
Finally in appendix A
the reader familiar with Mellin-Barnes representations will
find detailed instructions for using
Czakon's {\tt MB} program to extract the obstructions from an amplitude.

\section{Structure of the $L$-Loop Four-Gluon Amplitude}

We let $A^{(L)}(\epsilon,s,t)$ denote the planar $L$-loop four-gluon
amplitude in ${\cal N}=4$ super-Yang-Mills theory, dimensionally
regulated to $D = 4 - 2 \epsilon$ dimensions. We further define
\begin{equation}
M^{(L)}(x,\epsilon) = (s t)^{\epsilon L/2}
A^{(L)}(s,t,\epsilon)/A^{(0)}(s,t), \qquad x = t/s.
\end{equation}
It is convenient to divide by the tree-level amplitude because supersymmetry
determines the helicity structure of the $L$-loop amplitude to be the
same as that of the tree-level amplitude; therefore the ratio is a function
only of $\epsilon$ and the Mandelstam invariants.
The prefactor $(s t)^{\epsilon L/2}$ is convenient because
the resulting quantity $M^{(L)}$ is a function of the ratio $x = t/s$
only, and satisfies the symmetry
\begin{equation}
\label{eq:symmetry}
M^{(L)}(x,\epsilon) = M^{(L)}(1/x,\epsilon).
\end{equation}
However we caution the reader that it is not standard
in the literature
to factor out $(s t)^{\epsilon L/2}$ as we have done.

Expressions for $M^{(L)}(\epsilon, x)$ in terms of a basis of
integrals are known for
$L=1,2,3,4$~(\cite{Green:1982sw},
\cite{Bern:1997nh},
\cite{Bern:2005iz},
\cite{Bern:2006ew}
respectively).
For $L<4$ the basis is
determined by the structure of the simplest unitarity cuts, i.e,
double cuts. This analysis led to the ``rung rule" which basically
states that the $L$-loop amplitude can be constructed from the
diagrams that give rise to the $(L-1)$-loop amplitude by adding one
rung to all diagrams in such a way that no internal bubbles or
triangles are produced~\cite{Bern:1997nh}. For $L\ge 4$, the rung rule
gives only the part of the amplitude that can be determined by
double cuts. The four-loop amplitude is then special in the sense
that it is the first order in which the rung rule is incomplete.

In a remarkable recent effort~\cite{Bern:2006ew} the two diagrams
missing in the four-loop amplitude were found. There it was noticed
that these two diagrams, as well as the ones produced by the rung
rule, satisfy some intriguing conformality
properties~\cite{Kazakov:1984km,Usyukina:1993ch}
that have been
studied recently in~\cite{Drummond:2006rz}. This new insight,
together with some generalized unitarity arguments developed
in~\cite{Bern:1994cg,Bern:1994zx,Britto:2004nc,
Buchbinder:2005wp}, might lead to a
systematic way of generating the diagrams at any order in
perturbation theory. Very recently a proposal for the integrand of
the $L=5$ loop amplitude was presented~\cite{UCLAtalk}.
In Figure~1 we show the integrals contributing to the $L \le 4$ loop
amplitudes.

\begin{figure}
\begin{picture}(230,110)(0,0)
\Boxc(10,100)(20,20) \Boxc(10,70)(20,20) \Line(10,60)(10,80)
\Boxc(10,40)(20,20) \Line(6.6667,30)(6.6667,50)
\Line(13.3333,30)(13.3333,50) \Boxc(40,40)(20,20)
\Line(36.6667,30)(36.6667,50) \Line(36.6667,40)(50,40)
\Boxc(10,10)(20,20) \Line(5,0)(5,20) \Line(10,0)(10,20)
\Line(15,0)(15,20) \Boxc(40,10)(20,20) \Line(35,0)(35,20)
\Line(40,0)(40,20) \Line(40,10)(50,10) \Boxc(70,10)(20,20)
\Line(65,0)(65,20) \Line(65,6.6667)(80,6.6667)
\Line(72.5,6.6667)(72.5,20) \Boxc(100,10)(20,20) \Line(95,1)(95,20)
\Line(95,6.6667)(110,6.6667) \Line(95,13.3333)(110,13.3333)
\Boxc(130,10)(20,20) \Line(125,0)(125,20) \Line(135,0)(135,20)
\Line(125,10)(135,10) \Boxc(160,10)(20,20) \Line(160,0)(160,20)
\Line(150,6.6667)(160,6.6667) \Line(160,13.3333)(170,13.3333)
\Boxc(190,10)(20,20) \Line(180,0)(190,20) \Line(190,0)(200,20)
\Line(185,10)(195,10) \Boxc(220,10)(20,20) \Line(220,0)(220,20)
\Line(210,10)(230,10)
\end{picture}
\caption{Here we depict the topologies of the integrals which
contribute to the $L$-loop four-gluon amplitude, for $L \le 4$.  The
numerical coefficients and the numerator factors in the integrand
are given by the rung rule for all but the last two four-loop
diagrams.  We refer the reader to~\cite{Bern:2006ew} for
further details.} \end{figure}
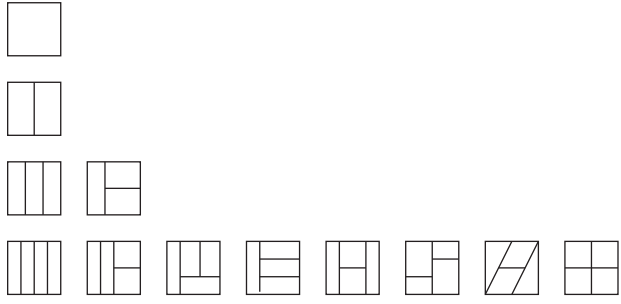

\subsection{Infrared Singularity Structure}

Let us now review the known IR singularity structure of
scattering amplitudes of gluons in ${\cal N}=4$ SYM.
The cusp anomalous dimension controls part of the
$1/\epsilon^2$ singularity and therefore it can be computed once the
amplitudes are known explicitly.

In particular, it is
known~\cite{Magnea:1990zb,Catani:1998bh,Sterman:2002qn,Bern:2005iz} that the
infrared singularities of an $L$-loop gluon amplitude, which contain
$1/\epsilon^{2L}$ and lower poles, can be expressed as a polynomial
in lower-loop amplitudes according to the formula
\begin{equation}
\label{milo} M^{(L)}(x,\epsilon) =X^{(L)}\left[M^{(l)}(x,\epsilon)
\right] + \frac{1}{4}
f^{(L)} M^{(1)}(x,L\epsilon) + {\cal O}(1/\epsilon)
\end{equation}
where the polynomials $X^{(L)}$ are conveniently encoded in the formula
\begin{equation}
\label{defx} X^{(L)}\left[M^{(l)}(x,\epsilon)
 \right] =
M^{(L)}(x,\epsilon) - \log \left. \left( 1+ \sum_{l=1}^{\infty}a^l
M^{(l)}(x,\epsilon)\right)\right|_{a^L\;{\rm term}}.
\end{equation}
The one-loop amplitude takes the form
\begin{equation}
M^{(1)}(x,L\epsilon) = -\frac{2}{L^2\epsilon^2} + {\cal O}\left( 1
\right).
\end{equation}
Therefore, (\ref{milo}) implies that
the $L$-loop cusp anomalous dimension is given by the
following deceivingly simple formula
\begin{equation}
%\label{simple}
\label{cuspi} f^{(L)} = - 2 L^2 \left[ M^{(L)}(x,\epsilon) -
X^{(L)}\left[M^{(l)}(x,\epsilon) \right] \right]_{1/\epsilon^2},
\end{equation}
where the subscript on the right instructs us to take the coefficient
of the $1/\epsilon^2$ singularity.

Note that $f^{(L)}$ is a number; therefore the right hand side of
(\ref{cuspi}) must be $x$ independent.
The calculation of~\cite{Bern:2006ew} does not make this $x$-independence
manifest, so $f^{(4)}$ was computed for various values of $x$ as a
consistency check, as mentioned in the introduction.
The cancellation of the $x$ dependence in (\ref{cuspi}) is precisely
the delicate step that we manage to avoid with our obstruction
technique as we explain in the next section.

Relations similar to (\ref{milo}) hold in any massless gauge theory,
but in ${\mathcal{N}} = 4$ Yang-Mills theory it is believed that the
stronger statement
\begin{equation}
\label{abdk}
M^{(L)}(x,\epsilon) =X^{(L)}\left[M^{(l)}(x,\epsilon)
\right] +\frac{1}{4} f^{(L)}(\epsilon) M^{(1)}(x,L\epsilon) + C^{(L)} +
{\cal O}(\epsilon)
\end{equation}
holds, where $C^{(L)}$ are numerical constants and $f^{(L)}(\epsilon)$
are functions of $\epsilon$ only, with $f^{(L)}(0) = f^{(L)}$ being
the $L$-loop cusp anomalous dimension. This relation was originally proven
for the two-loop four-particle amplitude
in~\cite{Anastasiou:2003kj}, and has been subsequently shown to hold
for the three-loop four-particle amplitude~\cite{Bern:2005iz} as
well as the two-loop five-particle amplitude~\cite{Cachazo:2006tj,
Bern:2006vw}.

Below we will make use of~(\ref{abdk}) at two and three loops, as
well as~(\ref{milo}) at four loops,
so we record here the explicit forms of these
relations,
\begin{eqnarray}
M^{(2)}(x,\epsilon)&=&\frac{1}{2}\left( M^{(1)}(x,\epsilon
)\right)^2 + \frac{1}{4} f^{(2)}(\epsilon) M^{(1)}(x,2 \epsilon) + C^{(2)}
+ {\cal O}(\epsilon),\cr
M^{(3)}(x,\epsilon) &=&-\frac{1}{3}\left( M^{(1)}(x,\epsilon
)\right)^3 + M^{(1)}(\epsilon, x) M^{(2)}(x,\epsilon)
+ \frac{1}{4} f^{(3)}(\epsilon) M^{(1)}(x,3 \epsilon) + C^{(3)}
+ {\cal O}(\epsilon),\cr
M^{(4)}(x,\epsilon)&=& \frac{1}{4}\left( M^{(1)}(x,\epsilon)\right)^4
- \left( M^{(1)}(x,\epsilon)\right)^2 M^{(2)}(x,\epsilon)
+ M^{(1)}(x,\epsilon)M^{(3)}(x,\epsilon)\cr
&&\qquad + \frac{1}{2}\left(
M^{(2)}(x,\epsilon)\right)^2
+ \frac{1}{4} f^{(4)} M^{(1)}(x,4 \epsilon) + {\cal O}(1/\epsilon),
\end{eqnarray}
where~\cite{Anastasiou:2003kj,Bern:2005iz}
\begin{eqnarray}
\label{abdkdata}
\frac{1}{4}
f^{(2)}(\epsilon)&=&- \zeta_2 - \zeta_3 \epsilon - \zeta_4 \epsilon^2
+ {\cal O}(\epsilon^3), \cr
C^{(2)} &=& - \frac{1}{2} \zeta_2^2,\cr
\frac{1}{4} f^{(3)}(\epsilon)&=& \frac{11}{2} \zeta_4 +
\epsilon(6 \zeta_5 + 5 \zeta_2 \zeta_3) + \epsilon^2 (c_1
\zeta_6 + c_2 \zeta_3^2) + {\cal O}(\epsilon^3),\cr
C^{(3)} &=& \left( \frac{341}{216} + \frac{2}{9} c_1\right)
\zeta_6 + \left( - \frac{17}{9} + \frac{2}{9} c_2 \right)
\zeta_3^2.
\end{eqnarray}
The quantities $c_1$ and $c_2$ are currently unknown, but they drop
out of~(\ref{abdk}). The quantities $f^{(L)}(0)$
agree with the original
calculations of the cusp anomalous dimension up to three loops
carried out in~\cite{Moch:2004pa,Vogt:2004mw,Kotikov:2004er}.

\section{Obstructions and Their Applications}

In this section we first refine the definition of obstructions given
in~\cite{Cachazo:2006mq}, clarifying
their more formal mathematical structure.
This allows us to give an all order in $\epsilon$ formula for the
one-loop obstruction. We then explain the systematic way of
implementing the extraction of obstructions at higher loops using as
a starting point the program ${\tt MB}$ by
Czakon~\cite{Czakon:2005rk}. Finally,
we give the formula for the
$L$-loop cusp anomalous dimension in terms of the obstructions at
$\le L$ loops.

\subsection{Obstructions---Theory}

The state-of-the-art technique for calculating $L$-loop integrals
makes use of Mellin-Barnes representations
\cite{Smirnov:1999gc,Tausk:1999vh}
(see~\cite{SmirnovBook} for a thorough introduction).
This technique leads us
to work not with an amplitude $M^{(L)}(x,\epsilon)$ but rather with
its Mellin transform $F^{(L)}(y,\epsilon)$.  These are related by
\begin{equation}
\label{eq:intone} M^{(L)}(x,\epsilon) = \int_{-i \infty}^{+i\infty}
\frac{dy}{2 \pi i}\ x^y\,F^{(L)}(y,\epsilon).
\end{equation}
It is fairly simple to write down an expression for
$F^{(L)}(y,\epsilon)$ using the Feynman rules for any desired
$L$-loop diagram. However these expressions are unwieldy because
they typically involve multi-dimensional integrals of ratios of
gamma functions. For example, the eight diagrams which contribute to
the four-loop amplitude require Mellin-Barnes representations of
dimensionalities from 10 to 14.

In the representation~(\ref{eq:intone}) one typically cannot expand around
$\epsilon = 0$ under the integral sign because the function $F$
has singularities which collide to pinch the integration contour
as $\epsilon \to 0$.
We isolate these singularities using the formula
\begin{equation}
\label{principaldef}
\lim_{\omega \to 0} \frac{1}{y \pm \omega} = {\cal P}\frac{1}{y} \pm
\pi \delta(y)
\end{equation}
and the generalization to higher singularities, given by taking
derivatives of this equation with respect to $y$. The formula is
missing some familiar factors of $i$ because the variable $y$ runs
along the imaginary axis, and hence we
take $\omega$ real\footnote{It is subtle to make sense of
$\delta$-functions of complex arguments, and we make no attempt to
do so here.  Rather, we emphasize that in the analysis of this
section we keep the $y$ contour {\it
precisely} on the imaginary axis, meaning that one could rewrite all
of the formulas in this section in terms of the real variable $z = -
i y$ if one preferred. In this case one would have to take
$\omega$ imaginary in~(\ref{principaldef})}.

Isolating all of the singularities at $y=0$ leads to
a formula of the form
\begin{equation}
\label{eq:inttwo}
M^{(L)}(x,\epsilon) =  \fpint{-i \infty}{+i \infty} {d y \over 2 \pi i}
\ x^y\,H^{(L)}(y,\epsilon) + P^{(L)}(\log x,\epsilon),
\end{equation}
where $\fpint{}{}$ denotes a principal value integral, and
we can now expand in $\epsilon$ under the integral sign.
Since the Mellin transform of $\frac{\partial^k}{\partial y^k}
\delta(y)$ is proportional
to $\log^k x$, the residual terms called $P^{(L)}$ in~(\ref{eq:inttwo})
are guaranteed
to be polynomial in $\log x$ (order by order in $\epsilon$).
These terms were called `obstructions'
in~\cite{Cachazo:2006mq}
because their presence indicates a failure to be able to assemble
everything under a common $y$ integral at $\epsilon=0$.
We see therefore that {\it obstructions correspond to $\delta(y)$-function
singularities in the Mellin transform of an amplitude.}
In fact, the symmetry~(\ref{eq:symmetry}) guarantees that $P^{(L)}(\log x,
\epsilon)$ will be an even polynomial, and it also guarantees
that $H^{(L)}(y,\epsilon)$ is an even function of $y$.

Let us illustrate this idea by calculating the obstruction to all
orders in $\epsilon$ for the one-loop amplitude. This result is
useful since the $1/\epsilon^2$ term of $M^{(L)}(x,\epsilon)$ is
controlled in part by the $\epsilon^{2L-4}$ term of
$M^{(1)}(x,\epsilon)$ as can be easily seen from (\ref{defx}).

In terms of the function
\begin{equation}
\label{eq:oneloop} F^{(1)}(y,\epsilon) = - \frac{1}{2} {e^{\epsilon
\gamma} \over \Gamma(-2 \epsilon)} \Gamma(1 +  \half \epsilon + y)
\Gamma^2(-\half \epsilon+y) \Gamma^2(-\half \epsilon - y)
\Gamma(1+\half \epsilon-y),
\end{equation}
the one-loop amplitude is given by
\begin{equation}
M^{(1)}(x,\epsilon) = \int_{-i \infty}^{+i \infty} {dy \over 2 \pi
i}\, x^y\, F^{(1)}(y,\epsilon), \qquad -2 < \epsilon < 0.
\end{equation}
As $\epsilon \to 0$, the two poles at $y = \pm \half \epsilon$
collide with the integration contour.  We can put those poles
directly on top of the integration contour, with a principal value
prescription, at the expense of subtracting off one-half of the
appropriate residues as in~(\ref{principalvalue}). Therefore we find
\begin{equation}
M^{(1)}(x,\epsilon) = \fpint{-i \infty}{+i \infty} {d y \over 2 \pi
i} \,x^y\,F^{(1)}(y,\epsilon) + \frac{1}{2} {\rm Res}_{y = + \epsilon/2}[
x^y F^{(1)}(y,\epsilon)] - \frac{1}{2} {\rm Res}_{y = -  \epsilon/2} [x^y
F^{(1)}(y,\epsilon)].
\end{equation}
The quantity $F^{(1)}(y,\epsilon)$ under the integral
is to be expanded in $\epsilon$ {\it inside} the integral. The
relative sign between the last two terms arises because the pole at
$y = + \epsilon/2$ approaches the integration contour  from the left
while the $y = - \epsilon/2$ pole approaches the contour from the
right.
The obstruction is the sum of these two residues.  Using the explicit
formula~(\ref{eq:oneloop}), we find that the one-loop obstruction to all
orders in $\epsilon$ is
\begin{equation}
\label{obstru}
P^{(1)}(\log x, \epsilon) = \frac{\hat{c}_\Gamma(\epsilon)}{\epsilon} \left[
(x^{+\epsilon/2} + x^{-\epsilon/2}) (\psi(1 + \epsilon) - 2
\psi(-\epsilon) + \psi(1)) + (x^{+\epsilon/2} - x^{-\epsilon/2}) \log
x \right],
\end{equation}
where $\psi(z) = \frac{d}{dz } \log \Gamma(z)$ and the prefactor contains
the constant
\begin{equation}
\hat{c}_\Gamma(\epsilon) = \frac{e^{\epsilon \gamma}}{2}
\frac{\Gamma(1+\epsilon) \Gamma^2(1-\epsilon)}{\Gamma(1 - 2 \epsilon)}.
\end{equation}
As expected based on our general arguments, the
$\epsilon$ expansion of the
obstruction~(\ref{obstru})
is an even polynomial in $\log x$.
We have also obtained an analytic
formula for the two-loop obstruction to all orders
in $\epsilon$, but the expression is quite lengthy and involves some remaining
Mellin-Barnes type integrals that we do not know how to perform
analytically at finite $\epsilon$.

\subsection{Obstructions---Implementation}

In the previous subsection we gave a formal definition of
obstructions and used it to compute the one-loop obstruction to all
orders in $\epsilon$. However, at higher loops the calculation
rapidly becomes overwhelming, and must be implemented in a computer
algebra system. Fortunately we can use Czakon's ${\tt MB}$
program which is designed to aid in the manipulation of
Mellin-Barnes integrals as the starting point of the implementation.
Let us now outline the steps we took in order to use ${\tt
MB}$ to calculate obstructions. This outline will probably only be
of interest to those with some familiarity with Mellin-Barnes
integrals, or with ${\tt MB}$ in particular.
A more concise outline of this procedure is given in the appendix.

We begin with a Mellin-Barnes representation for each
integral $I$ contributing to the desired amplitude,
\begin{equation}
\label{alistwo} I(x,\epsilon) = \int \frac{dz_1}{2\pi i} \cdots \int
\frac{dz_n}{2 \pi i} \
x^{f(z_i,\epsilon)} F(z_i,\epsilon),
\end{equation}
where $f(z_i,\epsilon)$ is a linear function of the $z_i$ and $\epsilon$,
and $F(z_i,\epsilon)$ is in general a complicated ratio of $\Gamma$ functions
whose arguments are linear in the $z_i$ and $\epsilon$ (a four-loop example
is given below).
The integration variables run from $-i \infty$ to $+i \infty$ along
contours that can be chosen to be parallel to the imaginary axis,
with appropriate real parts.

The representation (\ref{alistwo}) of $I(x,\epsilon)$ generically
is not valid in a neighborhood of $\epsilon =0$, rather one has to
start with a representation valid around some $\epsilon
=\epsilon_0<0$. One then analytically continues the formula to a
neighborhood of $\epsilon =0$, where one is free to develop the Laurent
expansion of the amplitude that will exhibit all of  the IR poles. In
the process of this analytic continuation one must cross poles of
the various $\Gamma$-functions in $F(z_i,\epsilon)$. The end
result is the sum of the original integral, now valid in a
neighborhood of $\epsilon = 0$, plus a sum of residue terms which
have fewer integration variables than the original integral.

Once this process is completed, the coefficient at any desired order
in $\epsilon$ is given as a sum of multiple integrals, most of
them containing a factor of $x^{f(z_i)}$ to a power that is linear
in the remaining integration variables. Some other
integrals do not depend on $x$, or their $x$ dependence is an overall
factor which is a power of $\log x$.

In each of the integrals which contains a factor of $x^{f(z_i)}$
we perform a linear change of variables $f(z_i) \to y$.
The next step is to take all integrals that contain an $x^y$ factor
and combine them under a single $y$ integral sign. Generically, this
is impossible because the integration contour for $y$ can be different
in each individual term.  For example, two typical contributions might
look like
\begin{equation}
I_1(x) = \int_{\beta_1-i\infty}^{\beta_1+i\infty} \frac{dy}{2\pi i}\ x^y\,
\Gamma(y)H_1(y), \qquad I_2(x) =
\int_{\beta_2-i\infty}^{\beta_2+i\infty} \frac{dy}{2\pi i}
\ x^y\,\Gamma(y)H_2(y)
\end{equation}
where $\beta_1 > 0> \beta_2$ are on different sides of $y=0$.
Therefore there is an obstruction at $y=0$ that prevents us from
combining these two integrals. The obstruction is a simple pole in
this case. Therefore, we can write
\begin{equation}
I_1(x)+I_2(x) = \int_{\beta_2-i\infty}^{\beta_2+i\infty} \frac{dy}{2 \pi i}
\ x^y\,
\Gamma(y) \left( H_1(y)+ H_2(y)\right) + \Delta
\end{equation}
where $\Delta$ is the obstruction, i.e, the contribution obtained
from the residue at $y=0$,
\begin{equation}
\Delta = \frac{1}{2 \pi i} \oint_{y=0} x^y\,\Gamma(y)H_1(y).
\end{equation}

As mentioned before, the presence of these obstructions that prevent
one from naively combining all terms under the same integral sign
were discovered in~\cite{Cachazo:2006mq}. In that paper, the goal
was to prove relations of the form (\ref{abdk}) as an identity under
the integral sign. Therefore, obstructions were undesired features of
the Mellin-Barnes representation and differential operators were
designed to annihilate them. Since our main goal in this
paper is to compute the
cusp anomalous dimension, it is precisely the obstructions that we
are interested in.

In~\cite{Cachazo:2006mq}, we were content to collect all terms under
a single with an arbitrary contour $\beta$.  However, as we have
already indicated in the previous subsection, it turns out that the
most natural and useful notion is to take $\beta = 0$ so that the
$y$ contour sits along the imaginary axis, with a principal value
prescription. If $\beta$ is infinitesimally small---which in
practice means that it is sufficiently small to ensure that there
are no more poles between the $y$ contour and the imaginary
axis---then the final step of pushing the $y$ contour to the
imaginary axis can be accomplished via
\begin{equation}
\label{principalvalue}
\int_{\beta - i \infty}^{\beta + i \infty} \frac{dy}{2 \pi i} \ x^y\,H(y)
= \fpint{-i \infty}{+i \infty} \frac{dy}{2 \pi i}
\ x^y\,H(y) + \frac{{\rm sign}(\beta)}{2}
{\rm Res}_{y = 0}\,[ x^y H(y)].
\end{equation}

The final step in calculating the full obstruction in any
amplitude $M(\epsilon,x)$ is to add up all of the
contributing integrals $I(\epsilon,x)$.  The above steps provide
a constructive proof that any amplitude (and moreover, each
individual integral $I(x,\epsilon)$
contributing to an amplitude) can be written in the
form
\begin{equation}
\label{decomp} M(x,\epsilon) = \fpint{-i\infty}{+i\infty} {dy \over
2 \pi i} \ x^y\,H(y,\epsilon) + P(\log x,\epsilon)
\end{equation}
for some $H(y,\epsilon)$ and $P(\log x,\epsilon)$.
Note that the sum over contributing integrals $I(x,\epsilon)$ imposes
that $M(\epsilon,x)$
satisfies the symmetry~(\ref{eq:symmetry}).
This ensures that the resulting $H(y,\epsilon)$, after all contributions
are added together, satisfies $H(y,\epsilon) = H(-y, \epsilon)$, and
that the obstruction $P(\log x,\epsilon)$ must be symmetric under
$x \leftrightarrow 1/x$, so it is an even polynomial in $\log x$.

\subsection{Cusp Anomalous Dimension From Obstructions}

Finally we explain how to exploit obstructions to simplify the
calculation of the cusp anomalous dimension using
equation~(\ref{cuspi}).
There are two essential properties of obstructions that enable
this calculation.

The first essential property is that obstructions satisfy a product
algebra---i.e. if $P_1$ and $P_2$ are the obstructions in two
amplitudes $M_1$ and $M_2$, then the obstruction in the product
$M_1(x,\epsilon) M_2(x,\epsilon)$ is simply the product of
obstructions $P_1(\log x,\epsilon) P_2(\log x,\epsilon)$.

This is easy to see in $y$-space, where multiplication of
$M_i(x,\epsilon)$ becomes convolution of the corresponding
$F_i(y,\epsilon)$. Delta functions (or derivatives of delta
functions) convolved with other delta functions still gives delta
functions. However, delta functions convolved with a smooth function
give a smooth function, as does of course a smooth function
convolved with a smooth function.

This fact allows us to calculate, order by order, the obstructions
in the 1,2,3, etc.~loop amplitudes once and for all, and then they
can subsequently be plugged into~(\ref{cuspi}) to study the cusp
anomalous dimension.
Furthermore,any polynomial iteration relation of the form~(\ref{abdk})
that is satisfied by multi-loop amplitudes must also be satisfied
by their corresponding obstructions.  This is just the statement
that we can take the Mellin transform of both sides of the equation
and truncate to the terms proportional to (derivatives of) delta
functions at $y = 0$.

The second essential property is that the cusp anomalous dimension
only receives contributions from obstructions, and in fact only from
the leading obstruction proportional to $\delta(y)$ (as opposed to
$\frac{\partial^k}{\partial y^k}
\delta(y)$ for $k > 0$). To prove this property,
let us recall formula (\ref{cuspi}) for the $L$-loop cusp anomalous
dimension
\begin{equation}
\label{cuspitwo} f^{(L)} = -2 L^2 \left[ M^{(L)}(x,\epsilon)
- X^{(L)}\left[M^{(l)}(x,\epsilon) \right] \right]_{1/\epsilon^2}.
\end{equation}
Use the decomposition (\ref{decomp}) to write
\begin{equation}
M^{(L)}(x,\epsilon) = \int_{-i\infty}^{+i\infty} {dy \over 2 \pi i}
\ x^y\, {\cal{P}} \left[H^{(L)}(y,\epsilon)\right] + P^{(L)}(\log x,\epsilon)
\end{equation}
and
\begin{equation}
X^{(L)}[M^{(l)}(x,\epsilon)] =\int_{-i\infty}^{+i\infty} {dy \over
2 \pi i} \  x^y\, {\cal P}\left[K^{(L)}(y,\epsilon)\right]
+ Q^{(L)}(\log x,\epsilon).
\end{equation}
for some $K$ and $Q$.
In fact the product algebra structure tells us that
$Q^{(L)}(\log x, \epsilon )$ is given in terms of the obstructions
in lower loop amplitudes by
\begin{equation}
\label{qoqi} Q^{(L)}(\log x, \epsilon ) = X^{(L)}[P^{(l)}(\log x,
\epsilon)].
\end{equation}
Here we write the principal value integral as the integral of the
principal part, with a conventional integral
along the imaginary $y$-axis instead of $\fpint{}{}$.
The reason is that we would like to collect all of the terms
in equation~(\ref{cuspitwo}) under the single $y$ integral
\begin{eqnarray}
\label{eq:five}
\int_{-i\infty}^{+i\infty} {dy \over
2 \pi i}\
x^y\, \delta(y) f^{(L)} &=&
- 2 L^2
\int_{-i\infty}^{+i\infty} {dy \over
2 \pi i}\ x^y \, {\cal P} \left[H^{(L)}(y,\epsilon) -
 K^{(L)}(y,\epsilon) \right]_{1/\epsilon^2}
\cr && -2 L^2 \int_{-i\infty}^{+i\infty} {dy \over 2 \pi i}\ x^y
\left[ P^{(L)}\left(\frac{\partial}{\partial y},\epsilon\right) -
Q^{(L)}\left(\frac{\partial}{\partial y},\epsilon\right)
\right]_{1/\epsilon^2} \!\!\!\!\!\!\!\! \delta(y).
\end{eqnarray}
Note that we use $\delta$-functions that contain an extra normalization
factor of $2 \pi i$ to compensate for our normalization of the $y$
integral.  The $\epsilon$ expansion is performed inside the $y$
integral, as explained in the previous section, and the $1/\epsilon^2$
coefficient of the term on the second line is a polynomial
in $\partial/\partial y$
acting on $\delta(y)$ inside the integral.

Because equation~(\ref{eq:five}) must hold for all $x$,
the integrand must satisfy the equation pointwise in $y$.
This means that
the principal parts and the $\delta$-function parts must separately
satisfy the equation, and in particular the coefficient of
$\frac{\partial^k}{\partial y^k} \delta(y)$ must satisfy~(\ref{eq:five})
separately for each $k$.
By reading off the coefficient of $\delta(y)$ on
both sides of~(\ref{eq:five}) we conclude that
\begin{equation}
f^{(L)} = -2 L^2\left[ P^{(L)}\left(0,\epsilon\right)
 - Q^{(L)}\left(0,\epsilon\right)
\right]_{1/\epsilon^2}.
\end{equation}
Finally, using (\ref{qoqi}) we arrive at our main formula for
computing the $L$-loop cusp anomalous dimension $f^{(L)}$ from the
obstructions of $l \le L$ loop amplitudes,
\begin{equation}
\label{eq:four}
f^{(L)} = -2 L^2\left[ P^{(L)}\left(0,\epsilon\right)
 - X^{(L)}[P^{(l)}( 0 ,
\epsilon)] \right]_{1/\epsilon^2}.
\end{equation}

\section{Analytic results at one, two and three loops}

In this section we illustrate our method by showing that an analytic
computation of $f^{(2)}$ and $f^{(3)}$ is possible without computing
any $x$-dependent integrals. In order words, none of the technology
available for evaluating Mellin-Barnes integrals in terms of
harmonic polylogarithms is needed. We also show that relations
like (\ref{abdk}) that are expected to hold for the full amplitude
must and do hold for the obstructions alone.

A careful analysis of the one-loop amplitude
was carried out in section III.A and the
formula~(\ref{obstru}) displays the obstructions in this
amplitude to all orders in $\epsilon$.
The computation of $f^{(L)}$ only requires the
expansion of $P^{(1)}(\log x,\epsilon)$ through order $\epsilon^{2L-4}$.
Since in this paper we are interested in $L\leq 4$ it is sufficient to
expand $P^{(1)}$ through order $\epsilon^4$:
\begin{eqnarray}
P^{(1)}(\log x, \epsilon)&=&- {2 \over \epsilon^2} + \left[{2
\pi^2 \over 3} + {\log^2 x \over 4} \right] + \epsilon \left[{17
\zeta_3 \over 3} \right] + \epsilon^2 \left[{41 \pi^4 \over 720} +
{\pi^2 \log^2 x \over 24} + {\log^4 x \over 64} \right]
\cr
&&+ \epsilon^3 \left[{67 \zeta_5 \over 5} - {59 \pi^2 \zeta_3
\over 36} - {11 \zeta_3 \log^2 x \over 24}  \right] \cr && +
\epsilon^4 \left[ - {\pi^6 \over 4320} - {70 \zeta_3^2 \over 9} -
{53 \pi^4 \log^2 x \over 5760} + {\log^6 x \over 4608} \right] +
{\cal O}(\epsilon^5).
\end{eqnarray}

We have also obtained an exact formula for the
obstructions in the two-loop amplitude.
However instead of displaying the complete complicated formula
we show its expansion through order $\epsilon^2$, which is sufficient
for the purpose of isolating its contribution to $f^{(4)}$:
\begin{eqnarray}
&&P^{(2)}(\log x, \epsilon) = {2 \over \epsilon^4} + {1 \over
\epsilon^2} \left[ -{5 \pi^2 \over 4} - {\log^2 x \over 2} \right] +
{1 \over \epsilon} \left[ - {65 \zeta_3 \over 6} \right]
\cr
&&\qquad\qquad + \left[ - {\pi^4 \over 90} + {\pi^2 \log^2 x \over 24}
\right] + \epsilon \left[
- {463 \zeta_5 \over 10} +
{77 \pi^2 \zeta_3 \over 12}  + {25 \zeta_3 \log^2 x \over 12}
\right]
\cr
&&\qquad\qquad + \epsilon^2 \left[- {1999 \pi^6 \over 30240}
+ {95 \zeta_3^2 \over 18} + {17 \pi^4 \log^2 x \over 720}
+ {\pi^2 \log^4 x \over 32} + {\log^6 x \over 144} \right] +
{\cal O}(\epsilon^3).
\end{eqnarray}

At three loops we have worked out the obstructions through
finite order, finding
\begin{eqnarray}
&&P^{(3)}(\log x , \epsilon) = - {4 \over 3 \epsilon^6} + {1
\over \epsilon^4} \left[ {7 \pi^2 \over 6} + {\log^2 x \over 2}
\right] + {1 \over \epsilon^3} \left[ {31 \zeta_3 \over 3} \right]
\cr
&&\qquad\qquad+ {1 \over \epsilon^2} \left[
- {161 \pi^4 \over 3240} - {7 \pi^2 \log^2x \over 48}
- {\log^4 x \over 32}
\right]\cr
&&\qquad\qquad
+ {1 \over \epsilon} \left[
{967 \zeta_5 \over 15} - {965 \pi^2 \zeta_3 \over 108}
- {25 \zeta_3 \log^2 x \over 8}
\right]
\cr
&&\qquad\qquad
+ \left[
{244261 \pi^6 \over 1632960}
+ {107 \zeta_3^2 \over 18}
- {7 \pi^4 \log^2 x \over 576}
- {13 \pi^2 \log^4 x \over 256}
- {3 \log^6 x \over 256}
\right] + {\cal O}(\epsilon).
\end{eqnarray}
It is a simple but delightful exercise to check that these three
expressions obey the two- and three-loop iterative relations
(\ref{abdk}), (\ref{abdkdata}). Although amusing, it is of course
guaranteed from our general
discussion of the properties of
obstructions in the previous section.

\section{Four-Loop Calculation}

In this section we explain the details of our numerical computation
of the four-loop cusp anomalous dimension using obstructions.

The first step is to find an expression for the four-loop amplitude in
terms of a small number of integrals. As mentioned in section II,
the rung rule does not give the full amplitude for $L>3$. Finding
the remaining contributions turns out to be a very hard problem.
Luckily, in~\cite{Bern:2006ew}, the new non-rung-rule
contributions were found and
the full answer passed nontrivial consistency checks like unitarity
and the correct IR singular behavior. Here we take the diagrammatic
formula of~\cite{Bern:2006ew}, shown in Figure 1, as our starting point.

The second step is to find Mellin-Barnes representations for each of
the integrals contributing to the amplitude. This is a
straightforward process but such representations are not unique.
Here we give as an illustration
a 10-fold Mellin-Barnes representation for
the four-loop ladder integral
(the tenth integral $\int \frac{dy}{2 \pi i}
 \,x^y$ is suppressed)
\begin{eqnarray}
F(y,\epsilon)&=&\frac{e^{4 \epsilon \gamma}
}{\Gamma(-2\epsilon)}
\int \prod_{i=1}^9 {dz_i\over 2 \pi i}\
\Gamma(y-2\epsilon)
       \Gamma(1 + 2 \epsilon - y)
       \Gamma(-1 - \epsilon - z_1) \Gamma(-z_1)
\Gamma(-z_2)
\cr
&&\times
 \Gamma(1 + z_1 + z_2)
       \Gamma(-1 - \epsilon - z_3) \Gamma(2 + \epsilon + z_1 + z_3)
\Gamma(1 - z_2 + z_3)
\cr
&&\times
       \Gamma(1 - z_1 - z_2 + z_4)
\Gamma(-2 - 2 \epsilon - z_5) \Gamma(z_1 - z_5)
\Gamma(-z_4 + z_5)
\cr
&&\times
\Gamma(-3 - 2 \epsilon + z_2 - z_3 - z_4 - z_6)
\Gamma(-z_6)
\Gamma(3 + \epsilon + z_3 + z_5 + z_6)
\cr
&&\times
\Gamma(-3 - 2 \epsilon - y - z_7)
       \Gamma(1 - z_4 + z_7)
\Gamma(-3 - 3 \epsilon - z_8) \Gamma(z_5 - z_8)
\cr
&&\times
       \Gamma(4 + 2 \epsilon + y + z_8) \Gamma(-z_7 + z_8)
\Gamma(-1 - \epsilon + z_4 + z_6 - z_7 - z_9)
\cr
&&\times
\Gamma(-3 - 2 \epsilon - y - z_8 - z_9)
       \Gamma(-z_9)
\Gamma(3 + \epsilon + y + z_7 + z_9)
\cr
&&\times
\Gamma(1 - z_5 - z_6 + z_8 + z_9) /
\cr
&&
(\Gamma(-1 - 3 \epsilon - z_1)
\Gamma(1 - z_2)
\Gamma(3 + \epsilon + z_1 + z_3)
       \Gamma(-2 - 4 \epsilon - z_5)
\cr
&&\times
\Gamma(1 - z_4 + z_5)
\Gamma(1 - z_6)
       \Gamma(-3 - 5 \epsilon - z_8) \Gamma(1 - z_7 + z_8) \Gamma(1 - z_9)).
\end{eqnarray}
The reason we display this one is that this representation has one
less integration variable than the one given in~\cite{Bern:2006ew},
reflecting the non-uniqueness of Mellin-Barnes representations.

The third step is to
extract the obstructions at order
$1/\epsilon^2$ from the Mellin-Barnes integrals. This step gives
rise to a list with a large number of integrals. All these integrals
are independent of $x$.

The fourth step is the simplification process. It turns out that a
large fraction of these obstruction integrals can be reduced to
integrals with a smaller number of integration variables by
applications of Barnes lemmas and their corollaries (see appendix D
of~\cite{SmirnovBook} for a nice collection of lemmas).

The result of this process is a formula for $f^{(4)}$ in terms of
$13$ 5-fold integrals, $131$ 4-folds and many more $3$-, $2$- and
$1$-folds. As an illustration we give the following ``analytic''
formula for the four-loop cusp anomalous dimension:
\begin{eqnarray}
- \frac{1}{32} f^{(4)} &=&
%6
\Bigg[ \frac{1}{64}\int_{C_5} \prod_{i=1}^5 \frac{dz_i}{2 \pi i}
   \frac{\Gamma (-{z_1}-{z_2}-2) \Gamma (-{z_2}) \Gamma
    ({z_1}-{z_3}+1) \Gamma (-{z_2}-{z_3}-2)}{
\Gamma
    (-{z_2}-{z_3}-1) \Gamma (1-{z_4}) \Gamma
    ({z_1}+{z_4}+3)}
\cr && \qquad \qquad \times \Gamma
    (-{z_3}) \Gamma (-{z_1}+{z_3}-{z_4}-1) \Gamma (-{z_4})
    \Gamma ({z_1}+{z_4}+2) \Gamma (-{z_3}-{z_5}-1)\cr
&& \qquad \qquad \times
 \Gamma
    (-{z_3}-{z_4}-{z_5}-1) \Gamma (-{z_5}) \Gamma
    ({z_2}+{z_3}+{z_5}+2) \Gamma ({z_3}+{z_4}+{z_5}+1)
\cr &&\qquad\qquad \times
    \Gamma ({z_1}+{z_2}+{z_3}+{z_4}+{z_5}+4)
\Bigg] + 12~{\rm similar~terms}\cr
%69
&&+ \Bigg[ \frac{1}{12} \int_{C_4} \prod_{i=1}^4 \frac{dz_i}{2 \pi
i}
   \frac{\Gamma (-{z_1}-1) \Gamma (-{z_1}-{z_2}-2) \Gamma
    (-{z_2}) \Gamma ({z_1}+{z_2}+2) \Gamma (-{z_3})}
{\Gamma
    (-{z_1}-{z_2}-1)}\cr
&&\qquad\qquad \times \Gamma
    ({z_2}+{z_3}+1) \Gamma
    (-{z_1}-{z_2}-{z_3}-{z_4}-1) \Gamma (-{z_4}) \Gamma
    ({z_1}+{z_4}+1)\cr
&&\qquad\qquad \times \Gamma ({z_3}+{z_4}) \psi
    (-{z_1}-{z_4}-1) \psi ({z_3}+{z_4}+1)
\Bigg] + 130~{\rm similar~terms} \cr
%5
&&+ \Bigg[\frac{5}{12} \int_{C_3} \prod_{i=1}^3 \frac{dz_i}{2 \pi i}
  \frac{\Gamma (-{z_1}-2) \Gamma ({z_1}+2) \Gamma (-{z_2}-2)
    \Gamma ({z_1}-{z_2}) \Gamma ({z_2}+2)}
{\Gamma ({z_1}+3)}\cr &&\qquad\qquad \times \Gamma
    (-{z_1}-{z_3}-1) \Gamma (-{z_2}-{z_3}-1) \Gamma
    ({z_2}+{z_3}+2) \Gamma ({z_1}+{z_2}+{z_3}+3)
\cr &&\qquad\qquad \times \psi (-{z_1}-{z_3}) \Bigg] + 1305~{\rm
similar~terms}\cr &&+ \Bigg[\frac{1}{16} \int_{C_2} \prod_{i=1}^2
{dz_i \over 2 \pi i}
  \Gamma (-{z_1})^2 \Gamma ({z_1}) \Gamma (1-{z_2}) \Gamma
    (-{z_1}-{z_2}+1) \Gamma ({z_1}+{z_2}-1)\cr
&&\qquad\qquad \times \Gamma
    ({z_1}+{z_2}) \psi'(1-{z_2}) \Bigg] + 9227~{\rm similar~terms}\cr
     &&+ \Bigg[ - \frac{57661}{27648} \int_{C_1} \frac{dz_1}{2 \pi i} \Gamma(-z_1-1) \Gamma(-z_1) \Gamma(z_1+1)^2
\psi'''(z_1+2) \Bigg]\cr &&\qquad\qquad\qquad \qquad\qquad
\qquad + 3719~{\rm similar~terms~}\cr
&&\qquad
{\rm + ~numbers}.
\end{eqnarray}
In this expression the integration contours for each variable are
straight lines parallel to the imaginary axis with real parts given
respectively by \begin{eqnarray} C_5 &=& \left[ -\frac{6}{5}, -
\frac{24}{25}, - \frac{9}{25}, -\frac{8}{25},
-\frac{4}{25}\right],\cr C_4 &=& \left[ - \frac{17}{25},
-\frac{12}{25}, - \frac{4}{25}, - \frac{3}{25} \right],\cr C_3 &=&
\left[ - \frac{7}{4}, - \frac{5}{4}, +\frac{1}{2} \right],\qquad C_2
=  \left[ - \frac{1}{4}, + \frac{3}{4}\right],\qquad C_1 = \left[ -
\frac{1}{2} \right]. \end{eqnarray}

The final step is the numerical evaluation of these integrals,
which we performed using {\sc CUBA}'s Cuhre algorithm~\cite{Hahn:2004fe} as
implemented by Czakon's {\tt MB} program. It is
important to mention that the main source of numerical error comes
from $5$ and $4$ folds.
The full list of integrals is
available from the authors upon request.

In the simplification process we implemented only the first and
second Barnes lemmas and some of their simplest corollaries. We
believe that a more detailed analysis will show that some major
simplifications are still possible and a fully analytic answer is
within reach. We believe that this is a very important problem but
we leave it for future work.

\section{Conclusions and Future Directions}

In this paper we clarified the nature of `obstructions' in $L$-loop
four-gluon amplitudes in ${\mathcal{N}} = 4$ Yang-Mills theory.
Their name comes from the fact that they are the terms which
obstruct writing an amplitude as an inverse Mellin
transform of a smooth function near $\epsilon = 0$.
These obstructions were shown to
possess all the information needed for the computation of the cusp
anomalous dimension at $L$-loop order.

We implemented a systematic way of extracting obstructions, thus
leading to a new method for computing the cusp anomalous dimension.
As an application of our method we recovered the known two- and
three-loop results already given in the
literature~\cite{Anastasiou:2003kj,Moch:2004pa,Vogt:2004mw,Kotikov:2004er,
Bern:2005iz}.

We also used the recently obtained~\cite{Bern:2006ew}
integrand for the four-loop four-gluon amplitude, given by the diagrams
shown in Figure 1,
to compute numerically the four-loop cusp anomalous dimension.
Our result is
\begin{equation}
\label{concu} f^{(4)} = -117.1789(2),
\end{equation}
where the quantity in parentheses denotes the uncertainty in the last
digit (as reported by the CUBA numerical integrator).

It would be highly desirable to find an analytic formula for
$f^{(4)}$. This is something which is definitely within reach and we
leave it for future work. However, we can try to see how far one can
get with (\ref{concu}) and some additional assumptions.
The first assumption one can make is the KLOV
transcendentality principle~\cite{Kotikov:2004er,Kotikov:2006ts}, which
assigns a degree of trascendentality $k$ to $\zeta_k$, $1$ to $\pi$
and zero to rational numbers. The degree of a monomial is computed
as the sum of the degree of each factor. The transcendentality
principle then states
that $f^{(L)}$ is homogeneous of degree $2 L-2$. This means that
$f^{(4)}$ must be a linear combination of $\pi^6$, $\zeta_3^2$, and
$\pi^3\zeta_3$ with rational coefficients.
Clearly, if we take this principle literally there is nothing we can
conclude from (\ref{concu}) for there would be an infinite number of
possibilities consistent with it.

Let us take a stronger form of the principle which was observed in
integrability studies~\cite{Eden:2006rx}. The idea is to allow
only
$\zeta_k$ numbers to appear in the monomials, and only
with {\it integer} coefficients with sign $(-1)^{L+1}$.
Then at four loops we would have the ansatz
\begin{equation}
f^{(4)} = -(n_1 \zeta_2^3 + n_2 \zeta_2 \zeta_4 + n_3
\zeta_6 +  n_4 \zeta_3^2 )
\end{equation}
with four positive integers $n_i$.
There is of course ambiguity in the choice of $(n_1,n_2,n_4)$ since
$7 \zeta_6 = 4 \zeta_2 \zeta_4$ and $5 \zeta_2 \zeta_4 = 2 \zeta_2^3$.
Then we can ask how many inequivalent choices $(n_1,n_2,n_3,n_4)$
are consistent with our result~(\ref{concu})
within error bars.
It turns out that there is only one choice, namely
\begin{equation}
- (4 \zeta_2^3 + 24 \zeta_2 \zeta_4 + 50 \zeta_6 + 4 \zeta_3^2)
\approx -117.1788285,
\end{equation}
which is the value conjectured in~\cite{Beisert:2006ez}.
The next nearest value
\begin{equation}
- (2 \zeta_2^3 + \zeta_2 \zeta_4 + \zeta_6 + 73 \zeta_3^2)
\approx -117.1801235
\end{equation}
is over six sigma away from our numerical data. For comparison, we
note that even with the best precision obtained
in~\cite{Bern:2006ew} there are more than 200 inequivalent choices
of $(n_1,n_2,n_3,n_4)$ within the error bars.

Another interesting problem for the future is to study the
feasibility of a higher loop calculation. As suggested
in~\cite{Beisert:2006ib}, if the integrability result holds to all
orders in perturbation theory, then there must be some hidden
structure in field theory that allows the same power of computation.
We hope that the theory of obstructions elaborated in this paper can
be refined even more so as to produce a technique applicable in
practice to all orders in perturbation theory. There are some
indications to be hopeful about this. One is the fact that we were
able to obtain a closed formula for the one loop obstruction to all
orders in~$\epsilon$ (see~\cite{Gehrmann:2005pd} for
similar all-orders quantities).
The second is that, assuming the transcendentality
principle, the anomalous dimension should only receive contribution
from the highest degree terms. The rest must cancel out. It turns out
that in our calculation many integrals gave rise to terms of the
``wrong" degree which later on cancel. It is conceivable that the
terms of wrong degree might be isolated and thrown away
systematically leaving a simple expression for the highest
degree terms---mirroring the way in which our method of obstructions isolates
and systematically throws away the $x$-dependent terms
which we also know must cancel. We also leave this fascinating question for the
future.

It is also very tempting to suggest that a five-loop calculation
might be feasible with our technique or a further refinement of it.
Something certain is that a five-loop computation of the anomalous
dimension using the IR structure of the five-loop
amplitude~\cite{UCLAtalk} and
then directly using {\tt MB}
to numerically evaluate both sides of~(\ref{eq:one})
is well
beyond our current computational power (the numerical computation of the
four-loop anomalous dimension presented here already required over
a CPU-month on 3 GHz Xeon computers).

It would be interesting to understand the connection between
obstructions and the converse mapping techniques employed
for example in \cite{Friot:2005cu} to find asymptotic expansions
of amplitudes, and to see whether any of the simplifications we
obtained can be applied to the direct calculation of the cusp
anomalous dimension from three-point functions (as in
\cite{Moch:2005id}).

Finally it would be interesting to investigate the applicability
of similar techniques to other theories, including possibly QCD.

\begin{acknowledgments}

MS and AV are grateful
to A.~Jevicki and D.~Kosower for interesting related
converations, to Z.~Bern for several comments on the manuscript,
and especially
to I.~Klebanov for encouragement to revisit
the four-loop amplitude after the complete basis of integrals became
known~\cite{Bern:2006ew}.
The authors gratefully acknowledge the use of computer
resources at the IAS
School of Natural Sciences, where this project was started in the
spring of 2006.
The research of FC at the Perimeter Institute is
supported in part by funds from NSERC of Canada and MEDT of Ontario.
The research of MS is supported by NSF grant PHY-0638520. Any
opinions, findings, and conclusions or recommendations expressed in
this material are those of the authors and do not necessarily
reflect the views of the National Science Foundation.

\end{acknowledgments}

\appendix

\section{Implementation Details}

Here we provide a concise outline explaining a straightforward
way to harness the power of Czakon's {\tt MB} program to
calculate the $L$-loop cusp anomalous dimension using the method
of obstructions as outlined in the main text of this paper.
Recall that to calculate the cusp anomalous dimension we first  need
to extract the numerical coefficient of $\delta(y)/\epsilon^2$ in
the Mellin transform of the $L$-loop amplitude.
This may be accomplished as follows:
\begin{enumerate}
\item{For each diagram that contributes to the $L$-loop amplitude,
find a valid Mellin-Barnes
representation and run {\tt MB} to read off the coefficient
of the $1/\epsilon^2$ pole.  The result is a list {\tt X} of
$x$-dependent
integrals valid at $\epsilon = 0$.  In each resulting term the $x$-dependence
can take four possible forms: (a) independent of $x$, (b) proportional to
a power of $\log x$, (c) $x$ to the power of a linear function of the
remaining integration variables, or (d) as in (c) but with a prefactor
of some power of $\log x$.}
\item{Throw away any integrals in {\tt X} that contain
any power of $\log x$---these terms cannot contribute to the
cusp anomalous dimension.}
\item{In the terms containing $x$ to some power, make a linear
change of variables to a new variable $y$ to write the $x$ dependence
in each term as $x^y$.  Of course, the location of the $y$ contour
will in general be different in each term.}
\item{Take each integral which contains $x^y$ and perform an analytic
continuation (by running the {\tt MB} program again) to shift
the $y$ contour infinitesimally close to $y=0$.  Here `infinitesimally
close' means, in practice, close enough to guarantee that there
are no remaining poles between the $y$ contour and the imaginary
$y$-axis.  This procedure produces a new list of $x$-dependent
integrals {\tt Y}.  The $x$-dependence in each
term is either a power of $\log x$, $x^y$, or independent of $x$.}
\item{Throw away any integrals in {\tt Y} that contain any power
of $\log x$.}
\item{Finally, for each term that still contains
$\int dy\, x^y$, push the $y$ contour onto the imaginary axis
using~(\ref{principalvalue}) and throw away
the principal part as well as any terms containing $\log x$.
The result is a list of $x$-independent
integrals {\tt Z}.}
\item{Add together the integrals {\tt Z} obtained from all of the
diagrams contributing to the $L$-loop amplitude.
The resulting set of integrals adds up to a number $s^{(L)}$.
  The interpretation
of this number is that it is the coefficient of $\frac{1}{\epsilon^2}
\delta(y)$ in
the Mellin transform of the $L$-loop amplitude $M^{(L)}(x,\epsilon)$.
For example, at four loops we find $s^{(4)} \approx -174.285193(6)$.
}
\end{enumerate}
The $L$-loop cusp anomalous dimension is then given in terms
of the number $s^{(L)}$ and the obstructions in the lower-loop
amplitudes according to the formula~(\ref{eq:four})
\begin{equation}
f^{(L)} = - 2 L^2 \left[ s^{(L)} -
X^{(L)}\left[P^{(l)}(0,\epsilon)\right]_{1/\epsilon^2} \right],
\end{equation}
where $X^{(L)}$ is the polynomial defined in~(\ref{milo}) above.

The full $L$-loop obstruction (including all of the $\log^k x$ terms)
can be computed without too much additional work.
One simply follows the above outline, order by order in $\epsilon$,
but without throwing away any $\log^k x$ terms.

The above procedure could actually be simplified somewhat by
making an appropriate change of variables to $y$ and choosing the
contour for $y$
to lie along the imaginary axis in the original Mellin-Barnes
representation, before running {\tt MB}.  One would then use
{\tt MB} to push $\epsilon$ infinitesimally close to $\epsilon = 0$,
and then finally use formula~(\ref{principalvalue})
for the final push to $\epsilon = 0$,
obtaining a principal value integral plus obstructions.
Note that this is how we chose to present the
calculation of~(\ref{obstru}).
Although conceptually simpler,
we did not employ this method at four
loops
because some experimentation revealed that this method seemed to
give rise to a larger number of terms than the above procedure.


\begin{thebibliography}{99}
\bibitem{UCLAtalk}
  Talk at the UCLA workshop ``Is $N = 8$ supergravity finite?'' by
  H.~Johansson.
\end{thebibliography}
\end{document}